# Homomorphic Sortition – Single Secret Leader Election for PoS Blockchains

LUCIANO FREITAS, LTCI, Télécom Paris, Institut Polytechnique de Paris, France
ANDREI TONKIKH, LTCI, Télécom Paris, Institut Polytechnique de Paris, France
ADDA-AKRAM BENDOUKHA, CEA LIST, Université Paris-Saclay, France
SARA TUCCI-PIERGIOVANNI, CEA LIST, Université Paris-Saclay, France
RENAUD SIRDEY, CEA LIST, Université Paris-Saclay, France
OANA STAN, CEA LIST, Université Paris-Saclay, France
PETR KUZNETSOV, LTCI, Télécom Paris, Institut Polytechnique de Paris, France

In a *single secret leader election* protocol (SSLE), one of the system participants is chosen and, unless it decides to reveal itself, no other participant can identify it. SSLE has a great potential in protecting blockchain consensus protocols against denial of service (DoS) attacks. However, all existing solutions either make strong synchrony assumptions or have *expiring registration*, meaning that they require elected processes to re-register themselves before they can be re-elected again. This, in turn, prohibits the use of these SSLE protocols to elect leaders in partially-synchronous consensus protocols as there may be long periods of network instability when no new blocks are decided and, thus, no new registrations (or re-registrations) are possible. In this paper, we propose *Homomorphic Sortition* – the first asynchronous SSLE protocol with non-expiring registration, making it the first solution compatible with partially-synchronous leader-based consensus protocols.

*Homomorphic Sortition* relies on Threshold Fully Homomorphic Encryption (ThFHE) and is tailored to proof-of-stake (PoS) blockchains, with several important optimizations with respect to prior proposals. In particular, unlike most existing SSLE protocols, it works with arbitrary stake distributions and does not require a user with multiple coins to be registered multiple times. Our protocol is highly *parallelizable* and can be run completely *off-chain* after setup.

Some blockchains require a sequence of rounds to have non-repeating leaders. We define a generalization of SSLE, called Secret Leader Permutation (SLP) in which the application can choose how many non-repeating leaders should be output in a sequence of rounds and we show how *Homomorphic Sortition* also solves this problem.

## 1 INTRODUCTION

**Consensus and blockchains.** Since the advent of Bitcoin [47], *blockchain* systems have grown to an entire field of study in computer science. At a high level, a blockchain is a tamper-proof ledger of blocks of data issued by the system members. As a baseline, the chain of data blocks keeps record of asset transfers between the participants [47], but often it offers more general functionality [23]. To reach agreement on the order in which data blocks appear in the chain, blockchains resort to the fundamental problem of *consensus*.

Consensus is one of the most studied problems in distributed computing. Mainstream consensus implementations follow the classical leader-based approach [26, 43], where a single process is selected to be the *leader* whose task is to *propose* a value and orchestrate the agreement.

Authors' addresses: Luciano Freitas, LTCI, Télécom Paris, Institut Polytechnique de Paris, France, lfreitas@telecom-paris.fr; Andrei Tonkikh, LTCI, Télécom Paris, Institut Polytechnique de Paris, France, tonkikh@telecom-paris.fr; Adda-Akram Bendoukha, CEA LIST, Université Paris-Saclay, France, adda-akram.bendoukha@cea.fr; Sara Tucci-Piergiovanni, CEA LIST, Université Paris-Saclay, France, sara.tucci@cea.fr; Renaud Sirdey, CEA LIST, Université Paris-Saclay, France, renaud.sirdey@cea.fr; Oana Stan, CEA LIST, Université Paris-Saclay, France, oana.stan@cea.fr; Petr Kuznetsov, LTCI, Télécom Paris, Institut Polytechnique de Paris, France, petr.kuznetsov@telecom-paris.fr.

Luciano Freitas, Andrei Tonkikh, Adda-Akram Bendoukha, Sara Tucci-Piergiovanni, Renaud Sirdey, Oana Stan, and Petr Kuznetsov

**Secret leader election.** Using a publicly known leader as a proposer of a new block opens the door to the *Denial of Service (DoS)* attack. The adversary can try bringing down the leader, which may slow down the consensus protocol, and thus the blockchain as a whole. This issue can be addressed in multiple ways.

One alternative is to use so-called *leaderless* consensus protocols, as done, for example, in Red Belly blockchain [30]. However, a vast majority of existing blockchains build atop their own consensus protocols, which rely on leaders.

Alternatively, following the approach of Algorand [38], one can use verifiable random functions (VRF), to select a set of participants as potential leaders and then pick one of them using a deterministic tiebreaker (e.g., the lowest VRF output). However, many consensus protocols require a single leader to be elected, and those that can manage multiple elected leaders require additional communication steps in order to filter out a single proposal in the end.

Thus, one may want to pick a single leader beforehand and *hide* its identity until it makes its proposal, and then it is too late to mount a DoS attack against it. This functionality is captured by the *single secret leader election* abstraction (SSLE) first formalized by Boneh et al. [14].

**Expiring registration.** In many existing SSLE protocols, elections are run sequentially and each leader needs to *re-register* after being elected. This action might include refreshing a secret or shuffling some common known list, but it will always require other processes to agree on which steps were taken to re-register, which means consensus is necessary in these situations.

In a synchronous setting, a consensus round with a correct leader always terminates, and thus this correct leader will always be able to use this instance of consensus to re-register. However, in partially synchronous systems, this is not always the case: even if a correct leader is elected, the consensus round may still fail to terminate due to unbounded message delays. Therefore, most existing SSLE solutions cannot be employed in partially synchronous leader-based consensus protocols.

**Our contributions.** We present an SSLE solution that we call *Homomorphic Sortition*, based on *Threshold Fully Homomorphic Encryption (ThFHE)*. Unlike existing SSLE protocols, our protocol can be instantiated in a purely *asynchronous* message passing system with *non-expiring registration*. Indeed, our protocol allows different instances of leader election to be run in parallel, independently of each other: the elected processes do not need to take any extra steps in order to continue participating in future rounds. Therefore, our solution perfectly matches partially synchronous consensus protocols.

Our protocol can be efficiently used in *proof-of-stake* (PoS) blockchains that rely on the assumption that the participants controlled by an adversary can only hold a minor fraction of the total stake, typically less than one third. *Homomorphic Sortition* fairly accounts for the stake distribution across the participants: the probability of a party being elected is proportional to its share of stake among the candidates. This is achieved without requiring users to register multiple times in the system, greatly improving scalability.

On the practical side, our solution was designed to benefit from multiprocessing machines by designing FHE circuits that can be run independently. Moreover, we tend to often apply the same operation to multiple data, allowing us to take leverage of platforms that support SIMD operations. Furthermore, once the setup phase is complete, the protocol does not require any information from consensus when deployed in a blockchain, which allows us to run several instances of our protocol in a pipeline and completely off-chain. By using precomputations, i.e., by electing the leaders in advance, the block creation latency can thus be made comparable to that of deterministic, insecure round-robin leader scheme.

Some blockchains require having a sequence of non-repeating leaders (e.g., Tezos based on Tenderbake [4], or Cosmos based on Tendermint [22]), which allows them to provide a property called *deterministic finality*, in which a new block is certain to be produced within a certain number of rounds after the network stabilizes. To address this challenge, we propose a generalization of SSLE—*Secret Leader Permutation* (SLP) which allows for secretly electing a fixed-length permutation of processes, weighted based on their stakes. Our *Homomorphic Sortition* protocol is designed to solve this generalized problem too.

**Roadmap.** In Section 2, we formally state the problems of SSLE and SLP. In Section 3, after reviewing handy cryptographic tools, we present our Homomorphic Sortition protocol. In Section 4, we prove its correctness, and in Section 5—discuss its complexity. Section 6 reviews the related work and Section 7 concludes the paper with a discussion of open questions. Appendix A discuss other works that also hide information about an elected leader in a consensus protocol, without guaranteeing a single leader, Appendix B presents an in depth discussion about possible circuits that can be used in our protocol and Appendix C proves the correctness of one of the presented circuits in particular.

## 2 PROBLEM STATEMENT

Consider a set of processes $\Pi = \{p_1, p_2, \ldots, p_n\}$ forming a committee with stakes given by an array $S$. Hence, the total stake of the committee is $s_t = \sum_{i \in [1..n]} S[i]$, and we assume that $s_f < s_t/2$ can be controlled by *Byzantine* processes that may exhibit arbitrary behavior. In particular, they may collaborate, trying to prevent *correct* (non-Byzantine) processes from being elected. To this end, they might omit information, send different messages to different processes (equivocate), share not well-formed data, etc.

The communication channels are *reliable*: messages exchanged by correct processes are eventually delivered. We assume, however, that the communication is *asynchronous*, i.e., communication delays may not be bounded.[1]

A *Single Secret Leader Election (SSLE)* protocol outputs a public *voucher* $v$ and $n$ private *proofs* $\pi_1, \ldots, \pi_n$ where, for each $i$, $\pi_i$ is only known by $p_i$ and no information about it is revealed to any other process or the adversary.

The protocol also defines a publicly known Boolean function *verify* that takes as inputs the index of a process, a proof and a voucher. A process $p_i$ is said to be *elected* if $p_i$ can *claim* the voucher $v$ with its proof $\pi_i$, i.e., $verify(i, \pi_i, v)$ returns TRUE.

An SSLE protocol must satisfy the following properties:

- **Uniqueness:** All correct processes output the same voucher $v$ and there exists unique $i \in \{1, \ldots, n\}$ such that $verify(i, \pi_i, v) = \text{TRUE}$. Moreover, the probability of the adversary producing a proof $\pi$ such that for $p_j \neq p_i$, $verify(j, \pi, v) = \text{TRUE}$ is negligible, even if it knows $\pi_i$.
- **Fairness:** The probability of process $p_i$ being elected is proportional to the amount of stake it holds in the committee. Formally, for all $p_i \in \Pi$:

$$\Pr[p_i \text{ is elected}] = \frac{S[i]}{\sum_{p_j \in \Pi} S[j]}$$

- **Unpredictability:** if the elected process is correct, unless the proof $\pi_i$ is revealed, the adversary cannot guess the leader with probability greater than $S[i]/(s_t - s_f)$.

---

[1]Notice that the consensus protocol that consumes the leader indications or the setup protocol may require stronger synchrony assumptions. In particular, most leader-based consensus protocols assume *partial synchrony*: after an unknown amount of time called *global stabilization time*, or *GST*), all the messages must arrive within a certain known upper bound time $\Delta$. Moreover, they typically assume a smaller threshold on the Byzantine stake: $s_f < s_t/3$ instead of $s_f < s_t/2$.

Luciano Freitas, Andrei Tonkikh, Adda-Akram Bendoukha, Sara Tucci-Piergiovanni, Renaud Sirdey, Oana Stan, and Petr Kuznetsov

- **Termination:** If all correct processes execute the protocol, then every correct process $p_i$ eventually locally outputs a proof $\pi_i$ and a voucher $v$.

The definition of *uniqueness* at the same time ensures the protocol elects a unique leader and also forbids the scenario where a leader reveals itself and a malicious process *steals* the election by communicating faster with other processes after seeing the elected process' proof. Meanwhile, the *fairness* property of the election translates into a fair distribution of blocks in the blockchain where the proportion of blocks produced by a user must be at least its proportion of stake in the system. As for *unpredictability*, it guarantees that the best strategy for the adversary to know the leader if it is not among corrupted processes is to use the stake distribution of the remaining processes to guess who the leader is. Finally, the termination property guarantees the progress of correct processes cannot be obstructed by Byzantine participants.

We also propose a generalization of SSLE which allows for electing a *sequence* of $d$ distinct leaders. We call this generalization of SSLE *Secret Leader Permutation (SLP)*, with SSLE being a special case of SLP where $d = 1$. An SLP protocol outputs a list of $d$ public *vouchers* $v_1, \ldots, v_d$ and $n$ private proofs $\pi_1, \ldots, \pi_n$ and satisfies the following properties:

- **Uniqueness:** $\forall r \in \{1, \ldots, d\}$, all processes output the same voucher $v_r$ and there exists unique $i \in \{1, \ldots, n\}$ such that $verify(i, \pi_i, v_r) = \text{TRUE}$. Moreover, the probability of the adversary producing a proof $\pi$ such that for $p_j \neq p_i$, $verify(j, \pi, v_r) = \text{TRUE}$ is negligible, even if they know $\pi_i$.
- **Fairness:** The leader in every round $r \in \{1, \ldots, d\}$, denoted $i_r$, is selected from $\Pi \setminus \{i_1, \ldots, i_{r-1}\}$ with probability proportional to its fraction in *not yet claimed* amount of stakes. Formally, for all $p_i \in \Pi$:

$$\Pr[i_r = p_i] = \begin{cases} 0, & \text{if } p_i \in \{i_1, \ldots, i_{r-1}\} \\ \frac{S[i]}{\sum_{p_j \in (\Pi \setminus \{i_1, \ldots, i_{r-1}\})} S[j]}, & \text{otherwise} \end{cases}$$

- **Unpredictability:** if the elected process $p_i$ in a given round $r$ is correct, unless the proof $\pi_i$ is revealed, the adversary cannot guess the leader for the round $r$ better than based on the information about the stakes, the set of already revealed leaders and the processes it has corrupted.
- **Termination:** If all correct processes execute the protocol, then every correct process $p_i$ eventually locally outputs the proof $\pi_i$ and vouchers $v_1, \ldots, v_d$.

Here we define the properties of SSLE for 1 round and *SLP* for $d$ rounds, but any number of rounds of consensus can be executed by repeatedly running the protocol, obtaining an ever growing sequence of leaders.

## 3 PROBLEM SOLUTION

Before presenting our Homomorphic Sortition protocol, we introduce our notation scheme and recall the cryptographic primitives we use.

Scalar values are denoted by lowercase Latin and Greek letters (e.g., $f$ or $\pi$), vectors by uppercase Latin letter (e.g., $L$), and sets are represented by capital Greek letters (e.g., $\Lambda$). Values that are homomorphically encrypted are equipped with an underline (e.g., $\underline{S}$). We typically use $b$ to denote binary values (even though they might be encoded in fields larger than 2, by having 0 and 1). Subscripts might be added to variables to give more context when necessary. A list of values enclosed in square brackets (e.g., $[x, y, z]$) denotes a column vector, and $(\cdot)_i$ denotes a value signed by process $p_i$.

## 3.1 Cryptographic primitives

**Encryption and signature schemes.** First of all, we need a digital signature scheme, hence every process $p_i$ has a pair of public and private signature keys. For setup, as it will be later discussed, it is also necessary to have a symmetric encryption scheme. We also need a Threshold Fully Homomorphic Encryption system (ThFHE) [15], which enables computations over encrypted data. Moreover, the scheme guarantees that a certain threshold stake of key holders (defined by a threshold value) are able to cooperate to decrypt the computation results.

| Method | Inputs | Outputs |
| --- | --- | --- |
| ThFHE.Enc | $\kappa_p, [x_1, x_2, \ldots, x_n]$ | $[\underline{x_1}, \underline{x_2}, \ldots, \underline{x_n}]$ |
| ThFHE.Eval | $C$, Inputs | Evaluation of $C$ applied to Inputs |
| ThFHE.PDec | $\kappa_i, \underline{x}$ | the $i$-th share of $x$ |
| ThFHE.Ver | $y, \underline{x}, j$ | 1 if $y$ is the $j$-th share of $x$, 0 otherwise |
| ThFHE.Dec | $[y_1, y_2, \ldots, y_s]$ (distinct valid shares of $x$ holding at least $s_f + 1$ stake) | $x$ |

Table 1. Methods exported by the ThFHE scheme

The methods provided by the ThFHE scheme are shown in Table 1. After a *setup* is complete, each process $p_i$ has access to a private key $\kappa_i$ used to partially decrypt a ciphertext encrypted with the joint public key $\kappa_p$ known to every process. An encryption operation is denoted *ThFHE.Enc* and a partial decryption operation is denoted *ThFHE.PDec*. Once processes holding at least $s_f + 1$ stake issue a partial decryption operation for a given encrypted value, it is possible to retrieve the corresponding clear-text value. Shares corresponding to less than $s_f + 1$ stake reveal no information about the encrypted value. Additionally, before decrypting the final value using the operation *ThFHE.Dec*, a process can check that the partial decryption it receives corresponds to an encrypted value by calling *ThFHE.Ver*.

**FHE schemes.** There are many different implementations of FHE. Notably, some of these constructions, such as FHEW [33] and TFHE [29] [2] encrypt the input bit-wise [3], meaning that operations can be done on each bit separately. Other constructions, such as BGV [21] and BFV [34] encrypt inputs word-wise, meaning that each cipher-text contains information of multiple data that is packed together, notably allowing operations to be done accross multiple data (SIMD). The schemes of the first type are able to execute binary circuits where the possible operations are *XOR*s and *AND*s; while the second type uses arithmetic circuits, performing modular addition and multiplication over some pre-defined prime number.

Some noise is added to the ciphertexts as operations are done on encrypted data, particularly when multiplication (including AND) operations are performed. If this noise becomes too large, values cannot be decrypted anymore and the data is forever lost. An operation called bootstrapping can then be performed, reducing the noise present in ciphertexts and potentially allowing an infinite number of operations to be executed.

Specifically, there is a very efficient algorithm for performing bootstrapping in TFHE, but no efficient solution is currently known for schemes such as BGV and BFV. The latter schemes avoid the problem by only considering circuits with bounded *multiplicative depth* (the maximum number of chained multiplication operations) and by choosing parameters that guarantee safe noise levels.

---

[2] Here, TFHE stands for Torus FHE [29]. To avoid confusion, we use a different notation (ThFHE) to denote Threshold FHE.
[3] Even though TFHE was initially described by Chilloti et al. as a bit-oriented FHE-scheme, many recent works investigate the possibility to encrypt integers, and the use of its functional-bootstrapping feature. In this work, we use TFHE as initially described in [29].

We assume ThFHE has the properties presented in [15], namely *compactness*, *correctness*, *robustness*, *semantic security*, *plaintext extractability*, and *simulation security*.

| Circuit | Inputs | Outputs |
|---|---|---|
| $C_<$ | $\underline{x}, [y_1, y_2, \ldots, y_n]$ | $[\underline{x < y_1}, \ldots, \underline{x < y_n}]$ |
| $C_{01}$ | $[\underline{b_1}, \ldots, \underline{b_n}]$ where $\underline{b_i} = \underline{b_1} \implies \forall j \geq i : \underline{b_j} = \underline{1}$ | $[\underline{b_1}, \underline{b_2 - b_1}, \ldots, \underline{b_n - b_{n-1}}]$ |
| $C_{Sel}$ | $[\underline{x_1}, \underline{x_2}, \ldots, \underline{x_n}], [\underline{b_1}, \underline{b_2}, \ldots, \underline{b_n}], \exists! i : \underline{b_i} = \underline{1} \land \forall j \neq i, \underline{b_j} = \underline{0}$ | $\underline{x_i}$ |
| $C_{PRF}$ | $\underline{k}, x$ | $\underline{PRF(k, x)}$ |
| $C_H$ | $\underline{x}$ | $\underline{H(x)}$ |

Table 2. SSLE circuits

We present our protocol in an agnostic manner, so it can be instantiated in any chosen FHE scheme. Operations over encrypted data are done via the method *ThFHE.Eval* using publicly known circuits presented in Table 2. In Appendix B we give a list of possible implementations of the circuits depending on the chosen type of the FHE scheme. The motivation for operations performed by each circuit will become clear to the reader in the next section, where we present the protocol and the circuits are put into context.

For SSLE, we need the circuits $C_<$ to compare an encrypted scalar with each of the values in a plaintext vector, obtaining a binary vector as a result; $C_{01}$ to find the first number equal to 1 in a vector of non-decreasing binary numbers; and $C_{Sel}$ to extract a value from an array in the only position equal to 1 in a binary vector.

The circuit $C_{PRF}$ receives as input a ciphertext of a key ($\underline{k}$) and a plaintext value ($x$) and outputs the ciphertext of the value of $PRF(k, x)$, where $PRF$ is a *pseudo-random function* with output domain $[0, 1, \ldots, \delta - 1]$, meaning that $PRF(k, \cdot)$ is indistinguishable from a random function for anyone who does not know the secret key [4] $k$. The circuit $C_H$ evaluates a hash function $H$. The property we require for this function is collision resistance: the probability that the adversary can find two values $x, y$, s.t. $x \neq y$ and $H(x) = H(y)$ is negligible. Collision-resistant hash functions also provide the *input hiding* property: given a large random number[5] $x$ and two different arbitrary values $y$ and $z$, $H(x||y)$ is statistically indistinguishable from $H(x||z)$ [18].

Table 3 provides the additional circuits required for solving SLP. Notice that contrary to $C_<$ which compares an encrypted value with a plaintext array, $C_\leq$ has an encrypted array as its input. The circuit $C_{scale}$ changes the domain of a uniform random number and $C_-$ subtracts a scalar from a vector in positions indicated by another binary vector.

### 3.2 The Homomorphic Sortition Protocol

**Overview.** Our *Homomorphic Sortition* protocol (presented in Algorithm 1) assumes, as global knowledge, a stake distribution $S$, an encrypted random seed $\underline{q}$ and a list $\underline{T}$ of encrypted random numbers, which we will call *tickets*. Also, every process $p_i$ gets a hold of its ticket $t_i$ (we discuss the

---
[4]Note that for a known $k$, the function is deterministic.
[5]It is precisely because we need this random number that we use the PRF function.

**Variables**

| | | | |
|---|---|---|---|
| $E$ | 1 in the position of elected process for round $r$, 0 otherwise | $m$ | unselected stake |
| $L$ | 0 in positions left to the elected process, 1 otherwise | $s_t$ | total system stake |
| $s_f$ | upper bound on the Byzantine stake | $n$ | number of participants |
| $T$ | processes' tickets | $S$ | processes' stakes |
| $s_r$ | stake of round $r$ leader | $t_r$ | ticket of round $r$ leader |
| $v_r$ | voucher of round $r$ leader | $i_r$ | ID of the round $r$ leader |
| $\Lambda$ | partial decryptions of result | $x$ | random number |
| $U$ | stake partial sum | $Z$ | scaled stake partial sum |
| $q$ | random seed | | |

Table 4. Homomorphic Sortition variables

**Algorithm 1** Homomorphic Sortition for $p_i$

1: Given $\underline{q}, S, \underline{T}$
2: **operation** sortition($r, d$)
3:     $\underline{x} := ThFHE.Eval(C_{PRF}, \underline{q}, r)$
4:     **if** $r - 1 \mod d = 0$ **then**
5:        $\underline{U} := \left[ \sum_{j \in [1]} \underline{S}[j], \ldots, \sum_{j \in [n]} \underline{S}[j] \right]$
6:        $\underline{m} := \underline{U}[n]$
7:        $\forall i \in [n] : \underline{Z}[i] := \underline{S}[i]\delta/\underline{m}$
8:        $\underline{L} := ThFHE.Eval(C_<, \underline{x}, \underline{Z})$
9:     **else** // Never executed in SSLE, only in SLP
10:        $\underline{U} := ThFHE.Eval(C_-, \underline{s_{r-1}}, \underline{U}, \underline{L})$
11:        $\underline{m} := \underline{U}[n]$
12:        $\underline{x} := ThFHE.Eval(C_{scale}, \underline{x}, \underline{m})$
13:        $\underline{L} := ThFHE.Eval(C_\leq, \underline{x}, \underline{U})$
14:     $\underline{E} := ThFHE.Eval(C_{01}, \underline{L})$
15:     $\underline{s_r} := ThFHE.Eval(C_{Sel}, \underline{E}, S)$
16:     $\underline{t_r} := ThFHE.Eval(C_{Sel}, \underline{E}, \underline{T})$
17:     $\underline{i_r} := ThFHE.Eval(C_{Sel}, \underline{E}, [1..n])$
18:     $\underline{\pi_r} := ThFHE.Eval(C_{PRF}, \underline{t_r}, r)$
19:     $\underline{v_r} := ThFHE.Eval(C_H, \underline{\pi_r}||\underline{i_r})$
20:     Send $\langle$ **PVoucher**, $ThFHE.PDec(\underline{v_r})$ $\rangle$ to all
21: **upon receiving** $\langle$**PVoucher**, $(\tilde{v}_r)_j\rangle$ from $p_j$
22:     **if** $ThFHE.Ver(\underline{v_r}, \tilde{v}_r) \wedge (\cdot)_j \notin \Lambda_r$ **then**
23:        $\Lambda_r := \Lambda_r \cup (\tilde{v}_r)_j$
24: **upon** $\sum_{(\cdot)_j \in \Lambda_r} S[j] \geq s_f + 1$
25:     $v_r := ThFHE.Dec(\Lambda_r)$

**Algorithm 2** Combining consensus with Homomorphic Sortition: process $p_i$

26: Let $\pi_{i,r} = PRF(t_i, r)$ and $v_{i,r} = H(\pi_{i,r}||i)$
27: **for** $r \in \{1, 2, 3, \ldots\}$ **do**
28:     **before** executing consensus$_r$
       // If $r \mod d \neq 0$, wait sortition for $r - 1$
29:        sortition($r, d$)
30:     **upon** $v_{i,r} = v_r$
31:        Propose block in consensus$_r$ with $\pi_{i,r}$
32:     **upon receiving** $\pi_j : v_r = H(\pi_j||j)$ from $p_j$
33:        Accept $p_j$'s proposal in consensus$_r$

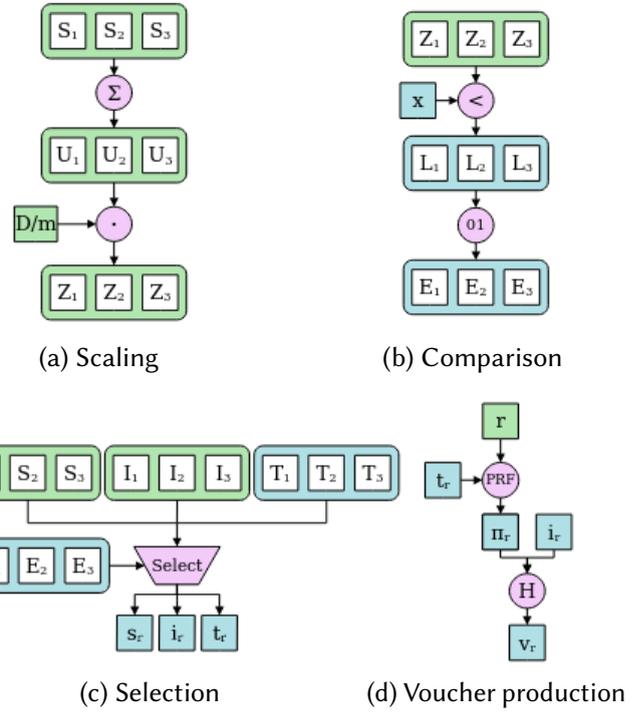

(a) Scaling      (b) Comparison

(c) Selection      (d) Voucher production

Fig. 1. SSLE components of Homomorphic Sortition

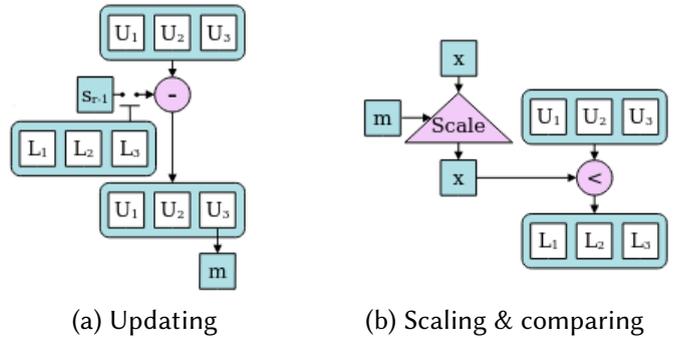

(a) Updating      (b) Scaling & comparing

Fig. 2. SLP components of Homomorphic Sortition

| Method | Inputs | Outputs |
|---|---|---|
| $C_\leq$ | $\underline{x}, [\underline{y_1}, \underline{y_2}, \ldots, \underline{y_n}]$ | $[\underline{x < y_1}, \ldots, \underline{x < y_n}]$ |
| $C_{scale}$ | $\underline{x}, m$ where x is a uniform random value over $\{0, \ldots, \delta - 1\}$ and $\delta \gg m$ | $\underline{y}$ uniformly distributed over $\{\underline{0}, \ldots, \underline{m-1}\}$ |
| $C_-$ | $\underline{y}, [\underline{x_1}, \ldots, \underline{x_n}], [\underline{b_1}, \ldots, \underline{b_n}]$ | $[\underline{x_1 - y \cdot b_1}, \ldots, \underline{x_n - y \cdot b_n}]$ |

Table 3. Additional circuits required for SLP

setup for this at the end of this section). An encrypted random number $\underline{x}$ is obtained via a PRF in the encrypted domain using the seed $\underline{q}$. This encrypted random number $\underline{x}$ is used, in conjunction with the stake distribution, to sample a member of the committee to be the leader in a fair manner with respect to the stake vector $S$. Each process then uses the encrypted version of the leader's ticket to produce a voucher representing the result of the election.

An instance of the protocol receives a round number $r$ and the length of the permutation $d$. If the application only requires SSLE, the instance is invoked with $d = 1$. To compute a sequence of leaders, one can repeatedly call the *sortition* operation with a incremented round number. Once a voucher is sampled, the corresponding process' information is updated in the sampling mechanism so that each user is selected at most once in a given sequence of $d$ SLP leaders. In both cases, once enough correct processes execute the protocol, the voucher is decrypted to its plain value (line 25).

Figure 1 shows the steps of the protocol corresponding to SSLE requirements and Figure 2 shows the steps corresponding to SLP.

**Scaling.** To elect the first leader in a permutation of SLP (or the only leader in SSLE), the protocol first builds an array of partial sums of the stakes, labeled $U$. The last element of this array is the total stake of the system, which at this point is also the total amount of *unselected* stake, denoted by $m$. The random number $x$ received as an input has a maximum value $\delta - 1$. We then scale up the partial sum array by multiplying every element by $\delta/m$ so the proportion of every element is maintained, but the last element of the new array, denoted by $Z$, becomes $\delta$.

**Comparison and leader selection.** By checking the first position of $Z$ that is greater or equal to $x$ we are able to pick a random process fairly. Intuitively, this is equivalent to having an interval partitioned in segments proportional to each process stake and then picking a point in the interval uniformly at random. To this end, we first check which indices of $Z$ store values that are strictly larger than $x$ (using circuit $C_<$) and then select the corresponding elements using circuit $C_{01}$, obtaining the array $E$ as a result (line 14). Note that before the comparison, all operations are done in clear data. Moreover, this comparison itself involves an encrypted number and clear data, which can be done much faster than comparing two encrypted numbers.

**Selection.** Once vector $E$ is built, it suffices to use the circuit $C_{Sel}$ to obtain the encrypted version of the leader's stake, ticket and the leader's identifier.

**Vouchers.** The goal of the sortition protocol is to determine who is the leader of a given round, however, this result must be announced in a manner that only the leader knows it was elected. Recall that each process $p_i$ holds a plain version of its ticket $t_i$ and an encrypted version of all tickets $\underline{T}$. While any elected process is capable of generating a proof to claim the winning voucher using the plain version of its ticket, the sortition protocol compute the winning voucher using the encrypted version of all tickets $\underline{T}$. Note that by not revealing its ticket, but only the proof, the tickets do not need to be refreshed and allows us to not have expiring registration.

The proof is the evaluation of the PRF with the winner's ticket and the round number. The properties of PRF ensure that disclosing the proof reveals nothing about the ticket and it is impossible to predict the elected process proof for any other round knowing its current proof (to be revealed later). Finally, the voucher is the hash of the winner's proof appended with the leader's identifier (in some implementations the id of the process might even replace the final bits of the proof, for making the hash evaluation more efficient). Hence, prior to the leader revealing itself, other processes cannot feasibly know who the leader is, as it would require inverting the hash function and breaking the PRF. Furthermore, once the leader reveals itself, producing a proof that would elect another process having access to this information would require breaking the collision resistance of the hash function.

**SLP: Updating.** Suppose now that the application needs a sequence of non-repeating leaders ($d > 1$). For $r \mod d \geq 1$, first, using $C_-$ and vector $L$, the stake of the previous leader $s_{r-1}$ is subtracted from the partial sum array $U$ in the position of the elected process, as well as in all positions to the right (line 10). Again, one can visualize the result as an interval of the length equal to the total unselected stake, where all segments to the left from the previous leader stay at their place, all segments to the right of are shifted by $s_{r-1}$. Notice that the amount of unselected stake will always be given by the last position of the partial sum array.

**SLP: Scaling and comparing.** Unlike the first round of SLP, to compute the $r$-th leader for $r \mod d \geq 1$, instead of scaling the partial sum array to the domain of the random number, we now to scale down the random number to the amount of unselected stake (lines 11 and 12). Hence, $L$ is now computed by comparing the shrunk random number with the unchanged partial sum array (line 13).

**Decryption.** The remaining part of Algorithm 1 describes how each voucher is decrypted (lines 20 to 25). Every process first broadcasts its partial decryption and waits until *valid* messages from the processes carrying $s_f + 1$ stake are received (a message is valid if it passes the partial decryption test). Once at least $s_f + 1$ stake is validated, the final decryption yields the plaintext value of the voucher.

**Combining consensus with Homomorphic Sortition.** One of the key advantages of our protocol is that once setup is done, it can be run asynchronously, *off-chain*, i.e., independently of the parallel blockchain consensus instances. However, the very goal of electing leader is precisely using them in the consensus protocol. Algorithm 2 describes how the leaders produced by our protocol can be fed to leader-based consensus instances. The result of homomorphic sortition for round $r$ must be obtained before executing the consensus instance for this round. A clever scheduling, combined with the pipelining of independent components of the protocol permit the result of a round to be always ready once it is time to execute the corresponding consensus instance, guaranteeing a global progress at the same pace as simply picking leaders in an insecure round robin fashion. Once a process is elected leader for round $r$, i.e., its proof claims the voucher of $r$, it proposes a block appending the proof. When a process receives a block and a proof for round $r$, it *accepts* the block if the proof is valid. Notice that SSLE rounds are completely independent and can be executed in any order, while in SLP, after the first round in a sequence, the other rounds can only be executed after the previous one finished. Different sequences remain independent, however.

**Setup.** Our protocol requires the ThFHE keys, the tickets and encrypted random seed to be determined at setup. For the first instance of the protocol, this might come in the form of a trusted setup containing all this information. Alternatively, one can rely on a consensus protocol (using round-robin leaders, for instance) for *Distributed Key Generation* (DKG), the tickets and the seed.

For example, every process can locally generate $n + 1$ random numbers and use consensus to agree on a set proposed by processes possessing at least $s_t - s_f$ stake. Finally, by adding (*xoring*) the vectors element-wise and using the first $n$ elements as tickets and the $(n+1)$-th element as the seed, we obtain a vector meeting the desired conditions. The seed is never decrypted, but each process must have access to its own ticket. This is achieved by symmetrically encrypting each ticket with a key only the holding process knows via FHE friendly block ciphers discussed in Appendix B, prior to decrypting them from the FHE domain. From time to time a new setup must be executed, as the distribution of stake will change and then it can be updated in the protocol, refreshing the encryption system. The good thing is that at this point, a new parallel consensus instance can be used for setup.

## 4 CORRECTNESS

Let $P$ be the space of proofs, $V$ be the space of vouchers, $R$ be the space of rounds.

We make use of the following game *steal* in Proposition 4.7.

(1) The challenger samples $i$ from $[n]$, following the probability distribution defined in the *fairness* property of SSLE, and then picks $t_1, t_2, \ldots, t_n$ at random from $T$ and $r \in R$. The challenger computes $\pi_i = PRF(t_i, r)$ and sends $H(\pi_i || i)$ and $r$ to the adversary.
(2) The adversary $\mathcal{A}$ picks a set of indices $J \subseteq [n]$, and sends it to the challenger.
(3) The challenger sends $\forall j \in J : t_j$ to $\mathcal{A}$.
(4) The adversary $\mathcal{A}$ submits a sequence of queries $(j_1, x_1), (j_2, x_2), \ldots$, where $\forall k : j_k \in [n]$ and $x_k \in R$.
(5) The challenger replies to each query $(j_k, x_k)$ with $PRF(t_{j_k}, x_k)$.
(6) The adversary $\mathcal{A}$ wins if for some $j \neq i$ it outputs $\pi_j \in P$ s.t. $H(\pi_j || j) = H(\pi_i || i)$.

We define the game *predict*, used in Proposition 4.9 by applying the following changes to *steal*:

- In step 4, the adversary cannot query the PRF evaluated in keys of processes it has not corrupted at the value $r$.
- In step 6, the adversary $\mathcal{A}$ outputs $j \in [n]$ and wins if $i = j$.

The following lemmata follow directly from the algorithm and can be easily checked to be true:

LEMMA 4.1. *The values in the partial sum array $U$ are increasing.*

LEMMA 4.2. $L[i] = 1 \implies \forall j \geq i, L[j] = 1$.

LEMMA 4.3. $i_r = i \implies \forall j < i, L[j] = 0$.

LEMMA 4.4. $L[n] = 1$

LEMMA 4.5. *No value is decrypted without at least one correct process issuing a partial decryption for it.*

PROPOSITION 4.6. *Homomorphic Sortition ensures termination.*

PROOF. The array $\underline{T}$ and the seed $\underline{q}$ are shared among all correct processes, as they are given by the setup, and the stake distribution is maintained common knowledge by the blockchain. Therefore, the participants will obtain the same values for the vouchers.

Moreover, the progress of the individual processes is only halted for waiting the partial decryptions of the vouchers (line 24). Since every round is executed by all correct processes and $s_f < s_t/2$, there will always be shares from sets of processes who hold at least $s_f + 1$ stake coming from correct processes executing the same instance. Hence, the decryption will always be successful and the protocol will always terminate. □

PROPOSITION 4.7. *Homomorphic Sortition ensures uniqueness.*

PROOF. First, let us show that in each round there is a unique voucher obtained by all correct processes.

*Homomorphic Sortition* is a protocol with deterministic steps, obtaining random results by having sources of randomness as its inputs. Notably, since the setup guarantees that every correct process has the same values for $\underline{T}$ and $\underline{q}$ and the stake distribution is public knowledge, all correct processes will issue decryptions for the same voucher. As our ThFHE scheme is *robust*, all correct participants decrypt the same voucher.

As the threshold system requires decryptions issued by participants holding at least $s_f + 1$ stake to produce a value, it is impossible for malicious processes to force the protocol to produce a second voucher, as it would imply some correct processes decrypt different vouchers, which we already ruled out.

Now, let us prove that in each round there exists a unique process that can claim the voucher. It is clear that at least one participant is able to claim the voucher because, by construction, our protocol selects the ticket of one of the participants and performs the same operations later needed to claim this voucher in order to produce it.

Let us show that the probability of any other member being able to claim this voucher is negligible. To this end, we will consider the attack game *steal* described in the beginning of this section. This game is a good model for breaking the uniqueness property of our protocol given that there is a unique winning voucher obtained by all correct processes, which we demonstrated earlier. Indeed, the *semantic security* of our ThFHE scheme ensures that the information available to the adversary is restricted to the secrets held by Byzantine processes and the data that is decrypted or exchanged in the clear. Hence, after the leader reveals itself, the information available to the adversary is restricted to the tickets of Byzantine participants and the proofs produced so far by the protocol (the knowledge of vouchers is redundant, given the proofs). Here, we allow the adversary to have even more freedom and query the proof of any round it wants in order to achieve its goal: to find another proof that claims the same voucher. Note that the adversary can know the proof for round $r$, and it should not be able to elect another process even after knowing this proof. The *advantage* of the adversary is then defined as the probability of it winning the game.

$$1adv[\mathcal{A}, \text{Sortition}] = \Pr[H(\pi_j || j) = H(PRF(t_i, r) || i)]$$

This probability is clearly negligible, as otherwise it would break the collision resistance of the hash function.

□

PROPOSITION 4.8. *Homomorphic Sortition ensures fairness.*

PROOF. The variable $m$ keeps track of the total unselected stake in the system. This invariant is maintained by initializing it with the total amount of stake $\sum_{p_j \in \Pi} S[j]$. In subsequent rounds, since $L[n]$ is always 1, the previous leader stake is always subtracted from $U[n]$, which is used to track $m$, maintaining the invariant.

The window of a process $p_j$ is the range of values $[U[j-1], U[j])$ for $j \in [2..n]$ and $[0, U[1])$ for $j = 1$. Initially, the window of every process $p_j$ has length $S[j]$, by construction. In the first round of a permutation, or in SSLE, every window size is multiplied by the same constant $\delta/m$, with the values of the array going from 0 to $\delta$. As a result, the interval $[0..\delta - 1]$ is split among all process, with process $j$ getting a fraction $S[j]/s_t$ of it. The leader is chosen by picking a point in the interval uniformly at random. Hence, the probability of process $i$ being elected in this case is:

$$\Pr[i_r = i] = \Pr\left[\left\lfloor \frac{\delta}{m} U[i-1] \right\rfloor \leq x < \left\lfloor \frac{\delta}{m} U[i] \right\rfloor\right] = \frac{\left\lfloor \frac{\delta}{m} U[i] \right\rfloor - \left\lfloor \frac{\delta}{m} U[i-1] \right\rfloor}{\delta}$$

Since we require $\delta \gg m$, this probability converges to $\frac{\frac{\delta}{m} \cdot S[i]}{\delta} = \frac{S[i]}{m}$, as required by the *fairness* criteria.

When running the protocol for SLP, after the election of process $p_i$, its window size becomes zero in the beginning of the next round, while the windows of other processes remain unchanged. This holds because the partial sum is updated by subtracting the elected stake $S[i]$ from every position where $L = 1$. Therefore, $U$ remains unchanged in every position less than $i$ (c.f. Lemma 4.3), keeping the corresponding processes windows unchanged, while for positions greater than $i$, both sides of the window are shifted by the same amount (c.f. lemma 4.2), conserving the length. However, at the round following the election of $p_i$, the lower end of its window is fixed, while the right is moved $S[i]$ to the left, so its length becomes $S[i] - S[i] = 0$.

The selection of a leader corresponds to the event where random variable $x$ falls within its window. Assuming that $x$ is a random number between 0 and $m - 1$, *fairness* follows immediately. □

PROPOSITION 4.9. *Homomorphic Sortition ensures unpredictability.*

PROOF. Let us consider the game *predict* presented in the beginning of this section. Differently than *uniqueness*, *unpredictability* only holds before the correct leader reveals itself, hence we must restrict the adversary to not evaluate the *PRF* at this point for processes it has not corrupted. The advantage of the adversary in this game, which we shall denote $Uadv$ is given by the probability it predicts process $i$ as the leader better than the information conveyed by the stakes and the corrupted processes:

$$Uadv[\mathcal{A}, \text{sortition}] = \begin{cases} \Pr[i = j] - \frac{S[i]}{s_t - \sum_{j' \in J} S[j']} & \text{if } i \notin J \\ 0 & \text{otherwise} \end{cases}$$

We can make use of the following hybrid arguments:

H0 the original game *predict*.
H1 replace the challenger's response in step 3 to $i$ if $i \in J$, $\bot$ otherwise.
H2 in step 1, the challenger picks $i$ from $[n]$ following the *fairness* distribution and produces proofs $\pi_1, \pi_2, \ldots, \pi_n$ at random by sampling $P$. The challenger then sends $H(\pi_i || i)$ to $\mathcal{A}$.
H3 delete step 4.
H4 the challenger only produces proof $\pi_i$.

H1 is indistinguishable from H0 because all tickets are independent, hence the only information obtained by the adversary is the identity of $i$, by testing the tickets it obtains if it manages to guess a set $J$ which contains it or the fact that $i \notin J$. H2 is indistinguishable from H1 because of the properties of the *PRF* function whenever the adversary does not know the key [6]. H3 is indistinguishable from H2 because the adversary is simply probing a true random function. H4 is indistinguishable from H3 because all other information which is deleted is never shared with the adversary, but kept locally. After applying these hybrids, we simplify the game to the following steps:

(1) The challenger selects $\pi \in P$ and $i \in [n]$ at random. It then sends $H(\pi || i)$ to $\mathcal{A}$.
(2) The adversary picks a set of indices $J \subseteq [n]$ and sends it to the challenger.

---
[6] A complete proof requires having the challenger to pick a random number $k \in \{0, 1, \ldots, n\}$, using PRF to generate proofs from numbers between 1 and $k$ and a true random function for the rest. Then, using standard proofs, it can be shown that the adversary can distinguish the results obtained between by $k = k'$ and $k = k' + 1$ with negligible probability. Finally, hybrid argument shows that the probability it can distinguish the case $k = 0$ and $k = n$ is also negligible

(3) The challenger replies the $\mathcal{A}$ with $i$ if $i \in J$, or $\perp$ otherwise.
(4) The adversary outputs $j \in [n]$ and wins if $j = i$.

If the adversary manages to obtain non-negligible advantage in this game, it means it obtains information about $i$ by analyzing the result of the hash function in the last game, otherwise the best it can do to find $i$ is to guess it correctly with probability $S[i]/(s_t - \sum_{j \in J} S[j])$, obtaining advantage 0. This contradicts the fact that our hash function provides the *input hiding* property (c.f. Appendix B), which informally implies that it is impossible to obtain any information about an input which is prefixed by a large random number. □

## 5 COMPLEXITY

### 5.1 Communication complexity

Let us first analyze the *Homomorphic Sortition* protocol from a distributed systems perspective, that is, by measuring the number of message round trips necessary per election (*time complexity*) and the number of words needed to be exchanged between processes (*communication complexity*). The only messages exchanged in the system are the partial decryption messages, which have all size $O(\lambda\beta)$, where $\lambda$ is the size of the hash function used and $\beta$ is the size of the encryption of a bit. In binary circuits, each bit is encrypted into $\lambda$ bits, while arithmetic circuits encrypt several integers at once, making $\beta$ a small number between 0 and 1. Therefore the latency of Homomorphic Sortition per round is of one message delay and the number of bits exchanged in total is $O(n^2\lambda\beta)$.

### 5.2 Computational complexity

For measuring the computational complexity of our protocol, it is necessary to know the complexity of the circuits it utilizes. Hence, we shall use the suggestions we present in Appendix B and analyze the number of gates used in a possible TFHE implementation that executes binary circuits and the multiplicative complexity of a possible BGV circuit executing our arithmetic circuits. These complexities are summarized in Table 5.

| Circuit | Nb of gates possible TFHE circuit | Multiplicative Depth possible BGV circuit |
| --- | --- | --- |
| $C_<$ | $O(n \log s_t)$ (2 inputs: one plain, the other encrypted) | 2 |
| $C_{01}$ | $O(n)$ | 1 |
| $C_{Sel}$ | $O(n^2)$ | 1 |
| $C_{PRF}$ | 4218 (AND gates only) | 6 |
| $C_H$ | 4218 (AND gates only) | 6 |
| $C_\leq$ | $O(n \log s_t)$ | $O(\log(n \log s_t))$ |
| $C_{scale}$ | $O((\log s_t)^2)$ | – |
| $C_-$ | $O(n)$ | 1 |

Table 5. Complexity of circuits

For more details on how each complexity was computed, please refer to Appendix B. We can conclude that our example TFHE solution would require the evaluation of $O(n^2)$ gates in both SSLE and SLP, while the suggested BGV implementation has multiplicative depth 16.

## 5.3 Threshold overhead

So far we have relied in FHE primitives for instantiating the circuits and anaylzing our complexity. In [16], authors develop a general approach for adding a threshold functionality that allows to split the secret key into a number $n$ of shares for LWE-based (Learning with errors) homomorphic schemes, with the property that the full decryption requires a number $t \leq n$ of partial decryptions (using secret key shares). The overhead due to a threshold scheme, as shown by the same work, comes from the set up phase and the decryption operations while the performances of homomorphic operations are unaffected by the "thresholdizing", so the operations themselves which are the most costly part are unchanged. The works of [12, 37] show how to obtain weighted forms of threshold systems required by our protocol.

## 6 RELATED WORK

Here we present an overview of the existing literature on SSLE protocols. As we shall see, most of earlier solutions are subject to expiring registration and/or do not account for stake distribution in an efficient way. In Appendix A, we discuss related work that, though not exactly in the context of SSLE, also concern with hiding information about the proposing process in some manner.

**Original SSLE formulation.** Boneh et al. [14] formally defined the problem of a single secret leader election (SSLE), [14] and described three way of solve SSLE: *Obfuscation* [8, 36], *Threshold Fully Homomorphic Encryption (ThFHE)* [15], and *Shuffling* [39]. The first technique was proposed to demonstrate theoretical feasibility of SSLE, while the other two were designed to be implemented in practice.

The ThFHE solution in [14] consists in generating an encrypted random number of $\log n$ bits and then expanding it to a vector of size $n$ where exactly one position is equals to 1 and all the rest is equals 0. (We believe that this implicitly restricts this solution to $n$ being powers of 2). Then, similarly to our solution, they use this vector to select a single secret, and the leader can later reveal some information to prove it. Contrary to ours, though, their solution has expiring registration: the way they reveal the secret does not allow it to be reused (it would be similar to our solution using tickets instead of vouchers to claim the election).

The shuffling solution in [14] consists in producing a list of DDH pairs, where each entry corresponds to a user, and only this user is able to open the commitment. Once a leader is elected, they replace their secret and shuffle the list, masking previous values first. This way other processes are still not able to determine which position the leader occupies in the list. Notice this action of replacing a secret and agreeing on the new shuffled list using this specific leader requires synchrony.

**Quantifying the gains of SSLE** In subsequent work, Azouvi and Cappelletti [5] performed an in-depth analysis of potential performance gains of using SSLE. They compared protocols that elect one leader on expectation (that is, some rounds might elect none or several leaders) which are known as Probabilistic Leader Election – PLE – with SSLE. They found that compared to PLE, SSLE reduces the latency of block creation by 25% when the blockchain is under a *private attack*. In this attack, the adversary creates its own private chain in parallel, not sharing it with the rest of miners and trying to seize the longest chain.

**Native Ethereum SSLE.** The ethereum foundation announced an upcoming SSLE protocol based on Boneh's solution with shuffling, called Whisk [3], in which they made some modifications which makes the shuffling routine more efficient, but does not guarantee perfect unpredictability as a trade-off. They have later announced a new version which is quantum secure [50].

**Using functional encryption.** Catalano, Fiore and Giunta [27] proposed a solution based on *functional encryption* [17, 48] in which, given a ciphertext $c$ encrypting a keyword $w$ and a secret key $sk$ associated to another keyword $w'$, the decryption allows one to learn if $w = w'$ and nothing more. The SSLE protocol of [27] is based on the idea that, for every election a small committee of users generates a ciphertext $c$ that encrypts a random keyword $j \in \{1, \cdots, N\}$, every user is given a secret key $sk_i$ associated with an integer $i$, and can claim victory by giving a non interactive zero knowledge (NIZK) proof that they can decrypt the election's ciphertext. They achieved a protocol that allows stake slashing of misbehaving nodes and, if executed without faults, the complexity of the protocol becomes sublinear on the number of participants. Although this protocol does not seem to have expiring registration, they do seem to require synchrony in order to produce unique ciphertexts.

**Adaptive secure SSLE.** Catalano, Fiore and Giunta [28] also modified Boneh et al.'s shuffling solution. In particular, they had the protocol maintain two lists: one which is kept in place, and is modified once the leader needs to re-register, while the other is shuffled. Moreover, they require the leader to produce an NIZK of the secret, instead of revealing the secret. After these changes, they prove the protocol to be adaptive secure and universally composable.

**Using MPC.** In [7], Backes, Berrang, Hanzlik and Pryvalov describe an MPC protocol where they explore the homomorphic properties of a secret sharing protocol to organize the processes in a tree structure. In the leaves of this tree, there are the secrets that will eventually elect one process as the leader, and they show how having access to a random beacon they are able to select one out of two secrets, and then applying this in their tree, to choose one out of $n$ secrets. Because of this structure, they show that considering weighted processes makes the latency of the protocol grow by $\log s_t$ where $s_t$ is the total stake of the system.

## 7 CONCLUDING REMARKS

We are currently working on implementing our SSLE/SLP solution using state-of-the-art ThFHE protocols, which will allow us to evaluate its performance in practice.

It would also be interesting to obtain an arithmetic implementation of the circuit $C_{scale}$ required for SLP, making it possible to use SIMD-friendly circuits for this scenario.

We also note that no existing SSLE solution, including ours, is able to produce correct results in a setting where the stake distribution among the committee members is highly dynamic. In alternative solutions requiring users to register multiple times, the protocol itself does not give such flexibility, as updating the secrets would be impractical. But in the case of our protocol, the encryption system becomes a bottleneck, which reflects the distribution of stake of keys at some point and requires the keys to be redistributed from time to time. The fact that our solution "freezes" the stake of processes that can be elected is compatible with many blockchains. Still, it would be interesting to either obtain a solution that operates with ever-changing stakes or to show such a solution is impractical for SSLE.

## A EXISTING NON SSLE PROTOCOLS THAT HIDE THE LEADER

**Algorand.** To the best of our knowledge, Gilad et al. [38] were the first to apply a secret election protocol to protect the consensus leader. In Algorand blockchain, two subsets of the users are selected: one to participate in the committee of users that will agree on which block should be appended and another (much smaller) to propose blocks – note that this is not a single leader election protocol. This selection is known as *cryptographic sortition* and it uses *verifiable random functions (VRFs)* [45] as follows: each user $i$ possesses a pair of public and private keys ($pk_i, sk_i$), respectively; each user $i$ can take the contents already published in the blockchain as a *seed* and apply a VRF to (*seed, $sk_i$*) generating a number $y$, which is indistinguishable from random and unpredictable for other users, and a proof $\pi$. Any other participant $j$ can then accurately verify that $y$ was generated with $VRF(seed, sk_i)$ only by analyzing *seed*, $\pi$ and $pk_i$ and without having to know $sk_i$. The system defines global thresholds so that each user whose random number complies with the verification is considered as selected. In the section, the protocol accounts the stake distribution so that the probability of selection is proportional to it and because the sortition is done locally, only the elected process knows its role before revealing it. Among potential leaders, ties are broken using a deterministic function (e.g., the minimal VRF output). Therefore, no participant can be sure that she is a leader before every participant reveals its output.

**Leader election as a smart contract.** In order to employ secret leader election in *Ethereum* [23], McCorry, and Meiklejohn [6] proposed a synchronous protocol that could be implemented by paying 0.10$ in gas per leader election once deployed in the targeted blockchain. The protocol advances in *rounds* from 1 to $round_{max}$, with each participant creating a sequence of hashes $(h_1, h_2, \cdots, h_{round_{max}})$ where $h_i = H(h_{i-1})$, publicly committing $H(h_{round_{max}})$. Every round *round* has a threshold $target_{round}$ and the processes generate together a common unpredictable random number $R_{round}$. Any process for which $(h_{round} \oplus R_{round}) < target_{round}$ is declared a leader of the round and can reveal its election by publishing $h_{round}$. In this system, any subset, including the empty set, of the participants can be elected; if the system requires that at most one process is elected, the processes can agree that the participant with the lowest local hash among the elected is the leader.

**Secrecy of stakes.** Ganesh et al. [35] formulated a variation on traditional Proof-of-Stake called *Private Proof-of-Stake (PPoS)* where the identities and stake of participants are not publicly available. Their work focuses on implementations based on VRF functions and shows how a participant can achieve the desired results by splitting its stake into several virtual accounts and by using an *anonymised* version of VRF (*AVRF*). In AVRF, proofs are still verifiable but have the property that it is extremely hard to tell whether two proofs have been created by the same secret key or not.

**Clusters of secrecy.** An interesting approach to the problem was suggested by Tan et al. [51]. In their system, the processes are organised in *clusters*, so that within a cluster, the identity of the leader is known, but processes outside it can only state which cluster contains the leader. In this manner, processes can choose trustworthy nodes they wish to form a cluster and form a resilient system using simple primitives relying only on local computations.

## B  CIRCUIT IMPLEMENTATIONS

$C_<$ takes a scalar $\underline{x}$ and a vector $\underline{Y}$ and outputs a binary vector where the $i$-th position is 1 if $x < y_i$, being 0 otherwise. [?] presents an efficient algorithm for both binary and arithmetic circuits which reduces the problem to a polynomial multiplication of a plain polynomial and an encrypted scalar. They further improve the solution for the binary-wise encryption case by avoiding unnecessary repeated comparisons when comparing multiple numbers, as in our case. Hence, the number of gates in the binary circuit corresponds to AND gates where one of the inputs is given by clear data and the multiplicative depth of the BGV circuit, as stated in the original paper, would be just 2.

$C_\leq$ takes a scalar $\underline{x}$ and (contrary to $C_<$) a vector $\underline{Y}$ and outputs a binary vector where the $i$-th position is 1 if $x < y_i$, being 0 otherwise. Binary circuits can efficiently achieve this by subtracting $\underline{x}$ from each element of $\underline{Y}$ and then taking the signal bit of each position, which can be efficiently done using the sign bootstrapping operation described in [20]; the complexity of this operation is hence the same as subtracting two numbers. An efficient circuit for doing comparison in BGV and BFV was presented in [42] whose multiplicative depth is proportional to the logarithm of the array size.

$C_{01}$ takes as an input a binary vector $\underline{B}$ which is increasing, in other words, after the position containing the first element equals 1, all subsequent elements are also 1. The output of this circuit is a vector which is 0 everywhere, except in the position of the first element 1. Equivalently, given a vector in the form $00 \ldots 011 \ldots 1$, the circuit determines the position of the transition between 0 to 1. This circuit can be implemented in binary schemes by the operation $\underline{B} \& ((\neg \underline{B}) \gg 1)$, i.e.: by performing an element-wise parallel AND between the vector itself and the rotation of its negation one position to the right (we consider the first element becomes zero after rotation). Arithmetic circuits can accomplish this by doing the operation $\underline{B} - (\underline{B} \text{ rot } 1) \cdot [011 \ldots 11])$, i.e.: subtracting the vector by itself cyclicacly rotated one position to the right, replacing the first position by zero after rotation.

$C_{Sel}$ takes a data vector $\underline{X}$, a binary vector $\underline{B}$ where one and only one position $i$ equals 1, while all others are 0 and outputs $x_i$. This is easily achieved in arithmetic circuits by making the dot product of the two vectors, while binary circuits simply replace the product operations of the arithmetic counterpart by ANDs.

The two following circuits rely on the evaluation of symmetric cryptographic primitives over FHE encrypted inputs (keys or message block). Many works have tackled this issue, mainly motivated by transciphering or proxy-reencryption [24, 25, 31, 32, 46] which aims at *compressing* the size of an FHE ciphertext given the ability to securely switch from a symmetric cryptosystem to an FHE one, by homomorphically decrypting the symmetric ciphertext. In other words, evaluate the symmetric decryption algorithm using an FHE encrypted key results in an FHE encryption of the message. Many *FHE-friendly* symmetric encryption schemes were designed with a focus on optimized decryption circuits for FHE operations. Namely MiMC [1], LowMC [2], Kreyvium [25], FLIP [46], and Rasta [31]. The decryption cirtcuit of these ciphers have a small multiplicative depth by design. Therefore, they are perfectly suited for an implementation of the solution using *levelled* FHE schemes as BGV or BFV. On the other hand, it turns out that TFHE performs better with lightweight ciphers which were initially designed for devices of small computational resources as discussed in [10]. Thereby, ciphers from NIST lightweight cryptography project [41] are good candidates when using gate-bootstrapping-based FHE schemes (FHEW / TFHE).

$C_{PRF}$ takes two arguments $k$ and $x$ and outputs $PRF(k, x)$, where $PRF : K \times X \to Y$ is a publicly known *pseudo-random function*, indistinguishable from a truly random function.[7] A *pseudo-random permutation* (PRP) is a restriction of a *PRF*, where $Y = X$, $\forall k \in K : f(k, \cdot)$ is one-to-one, and is also efficiently invertible. Bellare and Rogaway in [9] provided the PRP-PRF switching lemma which ensures that a *PRP* is indistinguishable from a *PRF*, and therefore, by transitivity, a secure *PRP* is indistinguishable from a truly random function.[8] In practice, the cryptographic primitive that models the abstract concept of a *PRP* is a secure block-cipher. Therefore, our $C_{PRF}$ cricuit is an evaluation of an FHE-friendly or lightweight block-cipher.

$C_H$ implements a cryptographic hash function $H$. *Iterated* hash functions in which the compression function contains a call or several calls to a block-cipher gave birth to provably secure constructions in the black-box model [13, 44, 49]. Thus, to design our secure FHE-friendly hash function that serves our protocol, we combine the following two paradigms: (1) the aforementioned FHE-friendly ciphers initially meant to be used for transciphering or other lightweight ciphers designed for low-resources environments, and (2) provably secure block-cipher-based constructions of iterated hash functions. Again for this circuit, the distinction between the binary mode using TFHE or FHEW or the arithmetic mode using BGV or BFV is done by selecting the proper cipher for each FHE class of schemes.

To implement a secure hash function that has good performance when run in the homomorphic domain we suggest an instantiation of Soichi Hirose's double-block-length hash construction from [40] with an ultra lightweight block cipher, namely PRINCE [19] for TFHE/FHWE implementations, or LowMC [2] for BGV/BFV implementations.

Hirose's construction provides optimal bounds for pre-image, and collision resistance in the black-box model. Formally, under the assumption that the underlying block-cipher is a random keyed invertible permutation, and that an adversary is given access to encryption and decryption oracles, finding a pre-image for Hirose's hash function requires at least $O(2^{2n})$ oracle queries, while a collision-finding attack takes an optimal adversary $O(2^n)$ where $n$ is the block length of the block-cipher. This block-cipher-based construction provides two advantages besides optimal security in the black-box model. The first one, is being a double-block length construction, that is, from an underlying block-cipher of block length $n$, it provides a hash function that outputs hashes of size $2n$. The second advantage is the fact that all double-block-length hash constructions with optimal security bounds make at least two calls to the block-cipher per round. Hirose's construction makes exactly two. In addition, the two calls are independent, and consequently executed in parallel. This provides a runtime comparable to single block length constructions from PGV [49] which make a single call to the block-cipher, and considerably better black-box security levels [11].

---

[7]We assume the the adversary is given $q$ oracle answers $(f(x_i))_i$ to its $q$ queries $x_i$, and decides whether $f$ is truly random or pseudo-random.

[8]The indistinguishability bound is $\frac{q(q-1)}{2^{\beta+1}}$ where $q$ is the number of oracle queries and $\beta$ is the size of the input/output.

The complexity of the $C_{PRF}$ circuit shown in Table 5 refers to the Rasta block cipher, operating with 256 bit long output. The $C_H$ circuit refers to the Rasta construction with 128 bit long output, being evaluated twice as part of the Hirose construction, which for this specific case gives the same values. Rasta was chosen to be presented in the table as they explicitly give the number of AND gates used in their protocol (for this same reason, these lines on the table only count and gates), while other constructions only mention their multiplicative depth.

$C_{scale}$ takes a uniformly distributed random number $x$ of $\beta_x$ bits between 0 and $2^{\beta_x} - 1$, as well as a number $m$ and outputs a uniformly distributed random number between 0 and $m - 1$. Considering that the number of bits $\beta_m$ used to represent $m$ is much smaller than $\beta_x$, binary circuits can implement this by multiplying the number $x$ by $m$ and then throwing away the $\beta_x$ less significant bits of the result.[9]

$C_-$ takes a scalar $y$ and two vectors $\underline{X} = [\underline{x_1}, \ldots, \underline{x_n}]$ and $\underline{B} = [\underline{b_1}, \ldots, \underline{b_n}]$, outputting a vector where the $i$-th position contains $\underline{x_i} - \underline{y}$ if $\underline{b_i} = 1$ or $\underline{b_i}$ otherwise. This can be accomplished in binary circuits by an AND operation between $\underline{y}$ and the vector $\underline{B}$, followed by a simple parallel element-wise subtraction. Arithmetic circuits can multiply $\underline{y}$ and $\underline{B}$, followed by an element-wise subtraction that is a SIMD operation.

## C  CORRECTNESS OF HOMORMORPHIC SORTITION USING PROPOSED $C_{scale}$

Let us show that our proposal of how to implement it using binary circuit in Appendix B is correct. The calculation of a scaled $x'$, i.e.: the multiplication of $x$ by $m$ and throwing away the $\beta_x$ less significant bits, corresponds to the expression $x' = \lfloor (xm)/\delta \rfloor$, because $\delta$ in this case is $2^{\beta_x} - 1 - (0) + 1 = 2^{\beta_x}$. Hence, the probability of a process $i$ being selected is given by:

$$P(i_r = i) = P(U[i-1] \leq x' < U[i]) = P\left(U[i-1] \leq \left\lfloor \frac{xm}{\delta} \right\rfloor < U[i]\right) =$$

$$P\left(U[i-1] \leq \frac{xm}{\delta} < U[i]\right) = P\left(U[i-1]\frac{\delta}{m} \leq x < U[i]\frac{\delta}{m}\right)$$

We can take the floor out of the probability, because the same interval given a fixed $x$ is described by it since stakes are integers. The probability of $i$ being elected, if it was elected before is therefore 0, as its window shall have size 0. On the other hand, if $i$ was never elected in the current permutation, its probability follow the same expression of the first round of SLP, thus, provided $\delta \gg m$, the election is always *fair*.

---

[9]As we shall see in the proofs, the distribution of each outcome of $y$ deviates from a uniform distribution by at most $2^{-(\beta_x - \beta_m)}$.